\input{aipcheck}

\documentclass[
    ,final            
  ]
  {aipproc}

\layoutstyle{6x9}
\usepackage{textcomp}

\begin{document}

\title{Coupling constants of the $S_{11}$ resonances to pseudoscalar mesons and octet baryons}


\author{C.~S.~An}{
  address={Institut de Recherche sur les lois Fondamentales de l'Univers, DSM/Irfu, CEA/Saclay,\\
F-91191 Gif-sur-Yvette, France}
}

\author{B.~Saghai}{
  address={Institut de Recherche sur les lois Fondamentales de l'Univers, DSM/Irfu, CEA/Saclay,\\
F-91191 Gif-sur-Yvette, France}
}

\begin{abstract}

Coupling constants of $S_{11}(1535)$
and $S_{11}(1650)$ resonances to meson-baryon ($\pi N$, $\eta N$, $\eta^\prime N$, $KY$)
are investigated in a chiral quark model, complemented by inclusion of  five-quark components in 
those resonances. 
The known strong decay partial widths of both resonances are well reproduced, provided
sizeable strangeness components in the relevant wave functions.
 
\end{abstract}

\maketitle


\section{Introduction}

Understanding the structure of the $S_{11}(1535)$ and $S_{11}(1650)$ resonances
and their dynamical properties are among important issues in baryon spectroscopy. 
At the hadronic level, $S_{11}(1535)$ is advocated to be a $K\Lambda-K\Sigma$ 
quasi-bound state~\cite{Kaiser:1995cy,Inoue:2001ip}, and in a recent paper~\cite{Bruns:2010sv}, 
$S_{11}(1650)$ is also reported to be a $K\Sigma$ bound state. 
In the quark models, both resonances are usually treated as the first orbitally excited states of 
the nucleon~\cite{Capstick:2000qj}. 
It is well-known that the structure of resonances can be investigated via photon- and hadron-induced 
production of mesons. 
So, the $S_{11}$-meson-baryon coupling constants are crucial issues in hadronic physics.

Recently, an Extended Chiral Constituent Quark Model (E$\chi$CQM) was developed to include higher Fock 
components in the resonances wave functions. Within that approach, the electromagnetic and strong decays 
properties of the $P_{31}(1232)$~\cite{Li:2005jn},
$P_{11}(1440)$~\cite{Li:2006nm}, $S_{11}(1535)$~\cite{An:2008xk} and 
$S_{01}(1405)$~\cite{An:2010wb,An:2010tv} resonances are well described.

Here, we present our recent calculations on the couplings of $S_{11}(1535)$ and 
$S_{11}(1650)$ to pseudoscalar mesons and octet baryons in the E$\chi$CQM. In our model~\cite{An:2011}, 
both resonances are considered as admixtures of three-quark
and strangeness five-quark components, consistent with the proposed $KY$ bound-state structures at 
hadronic level.

\section{Theoretical framework}

In this work, we employ the traditional wave functions~\cite{Capstick:2000qj}
for the three-quark components of $S_{11}(1535)$ and $S_{11}(1650)$. For
the five-quark components, explicit wave functions can be found in Ref.~\cite{An:2008xk}.
Here we will follow the formalism for strong decays in Ref.~\cite{An:2010wb}, where 
the coupling of a resonance to meson-baryon channel is obtained
by calculating the matrix element of the following operators:
\begin{eqnarray}
H^{NR(3)}_{M}&=&\sum_{j}\frac{g^{q}_{A}}{2f_{M}}(\frac{\omega_{M}}{E_{f}+M_{f}}\sigma\cdot\vec{P}_{f}+
\frac{\omega_{M}}{E_{i}+M_{i}}\sigma\cdot\vec{P}_{i}-\sigma\cdot\vec{k}_{M}
+\frac{\omega_{M}}{2\mu}\sigma\cdot\vec{p}_{j})X^{j}_{M}
\exp\{-i\vec{k}_{M}\cdot\vec{r}_{j}\},\nonumber\\
 H_{M}^{NR(5)}&=&\sum_{j}\frac{g^{q}_{A}}{2f_{M}}C_{XFSC}^{j}(m_{i}+m_{f})
 \bar{\chi}^{\dagger}_{z}\pmatrix{ 1&0\cr 0&1
\cr}\chi_{z}^{j}X_{M}^{j}\exp\{-i\vec{k}_{M}\cdot\vec{r}_{j}\}\,,
\label{op5}
\end{eqnarray}
with the meson emission operators being
\begin{eqnarray}
& X^{j}_{\pi^{0}}=\lambda_{3}^{j}, ~X^{j}_{\pi^{\pm}}=\mp\frac{1}{\sqrt{2}}(\lambda_{1}^{j}\mp\lambda_{2}^{j}),\nonumber\\
&
X^{j}_{K^{\pm}}=\mp\frac{1}{\sqrt{2}}(\lambda_{4}^{j}\mp\lambda_{5}^{j}),~X^{j}_{K^{0}}=
\mp\frac{1}{\sqrt{2}}(\lambda_{6}^{j}\mp\lambda_{7}^{j}),\\
&
X^{j}_{\eta}=cos\theta\lambda_{8}^{j}-sin\theta\sqrt{\frac{2}{3}}\mathcal{I},
~X^{j}_{\eta^\prime}=sin\theta\lambda_{8}^{j}+cos\theta\sqrt{\frac{2}{3}}\mathcal{I}\nonumber\,,
\end{eqnarray}
where $\lambda^{j}_{i}$ are the $SU(3)$ Gell-Mann matrices, and $\mathcal{I}$ the
unit operator in the $SU(3)$ flavor space. $\theta$ denotes the mixing angle between
$\eta_{1}$ and $\eta_{8}$, which leads to the physical states for $\eta$ and $\eta^{\prime}$.

\section{Numerical results}

The input parameters in our model are: 
constituent quark masses ($m$ and $m_s$), oscillator parameters ($\omega_{3}$ and $\omega_{5}$), 
mixing angle $\theta_{S}$ between $N^{2}_{8}P_{M}$ and $N^{4}_{8}P_{M}$ states, 
probabilities for five-quark components in  $S_{11}(1535)$ ($P_{5q}$) and $S_{11}(1650)$ ($P^\prime_{5q}$).
In line with Refs.~\cite{An:2008xk,An:2010wb}, we take 
$m=290$ MeV, $m_{s}=430$ MeV, $\omega_{3}=340$ MeV and $\omega_{5}=$ 600 MeV.

In order to calculate partial decay widths, for the remaining three adjustable parameters 
($\theta_{S}$, $P_{5q}$ and $P^\prime_{5q}$) the whole phase space was mapped out in the
following ranges: $0^\circ \leq \theta_{S}\leq 90^\circ$, 
$0\% \leq P_{5q} \leq 100\% $ and $0\% \leq P^\prime_{5q} \leq 100\%$.

Then, ranges were determined allowing to reproduce the known~\cite{Nakamura:2010zzi} partial decay 
widths for 
$S_{11}(1535)\to \pi N,~\eta N$, and $S_{11}(1650) \to \pi N,~\eta N,~K\Lambda$, and turned out to be
$26.9^\circ \leq \theta_{S}\leq 29.8^\circ$, 
$21\% \leq P_{5q} \leq 31\% $ and 
$11\% \leq P^\prime_{5q} \leq 18\%$.

\begin{figure}[t]
  \includegraphics[height=.32\textheight]{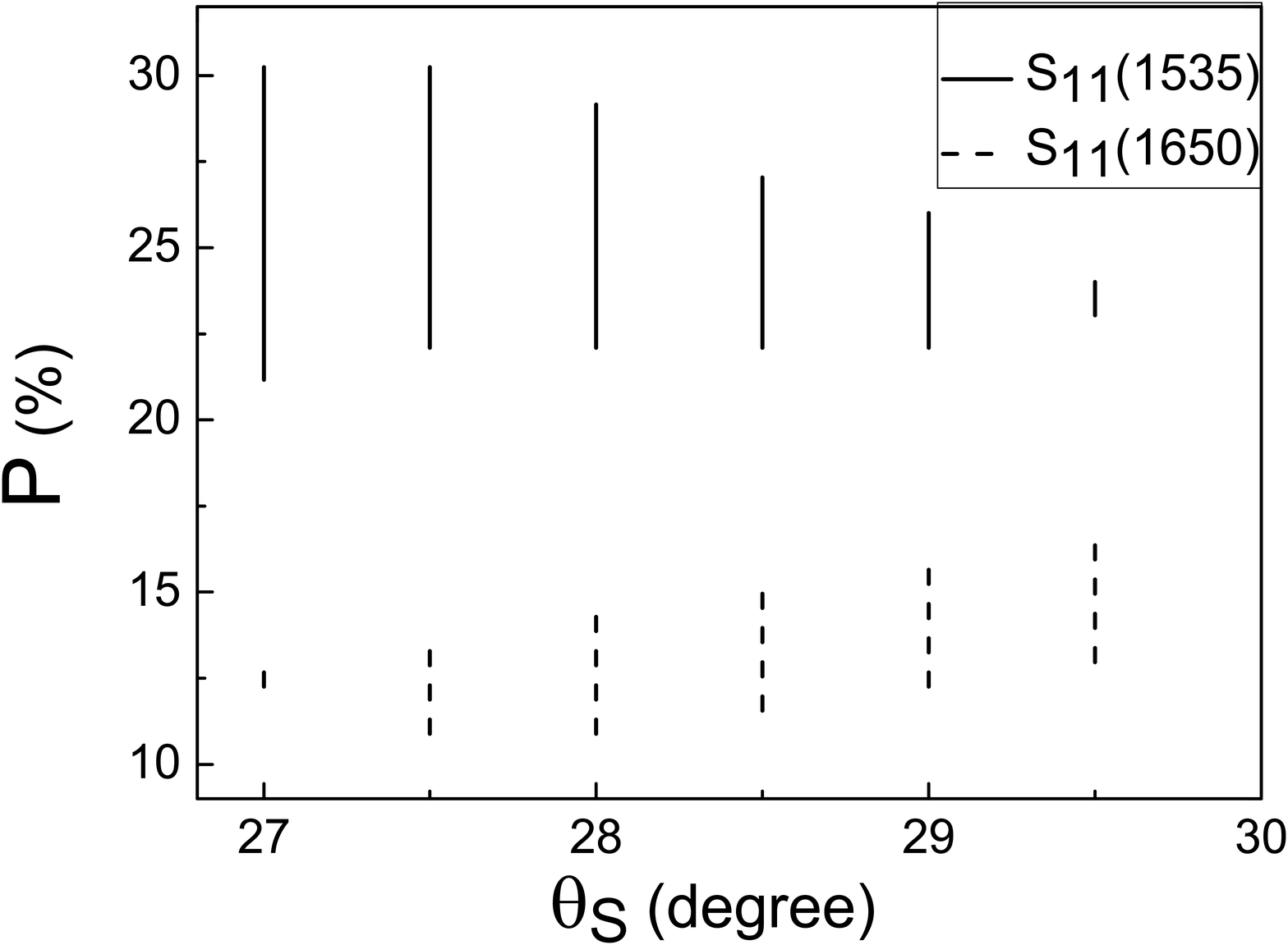}
  \caption{Ranges for five-quark components probabilities in $S_{11}$ 
as a function of mixing angle. }
\label{theta}
\end{figure}

In Fig.~\ref{theta} the relevant ranges for the five-quark probabilities at various mixing angles
are depicted, showing smooth but significant angle dependence.
Within that subspace, extreme values for the partial widths have been extracted  
(Table~\ref{width}), and compared to the data, resulting in from good to excellent agreements.
\begin{table}[h]
\begin{tabular}{lccccccccccc}
\hline 
$N^*$          && $\pi N$ && $\eta N$&& $K \Lambda$ && Ref.   \\
\hline
$S_{11}(1535)$ && 68 $\pm$ 15 && 79 $\pm$ 11 &&   &&                 PDG~\cite{Nakamura:2010zzi}  \\
               && 58 $\pm$ 5 && 79 $\pm$ 11 &&       &&  Present work \\        
$S_{11}(1650)$ &&128 $\pm$ 29&& 3.8 $\pm$ 3.6 && 4.8 $\pm$  0.7 &&          PDG~\cite{Nakamura:2010zzi} \\
               &&143 $\pm$ 5 && 4.5 $\pm$ 2.8 && 4.5 $\pm$ 0.5  &&  Present work \\
              
\hline
\end{tabular}
\caption{Strong decay widths for $S_{11}(1535)$ and $S_{11}(1650)$.}
\label{width}
\end{table}

\begin{figure}[hb]
  \includegraphics[height=.42\textheight]{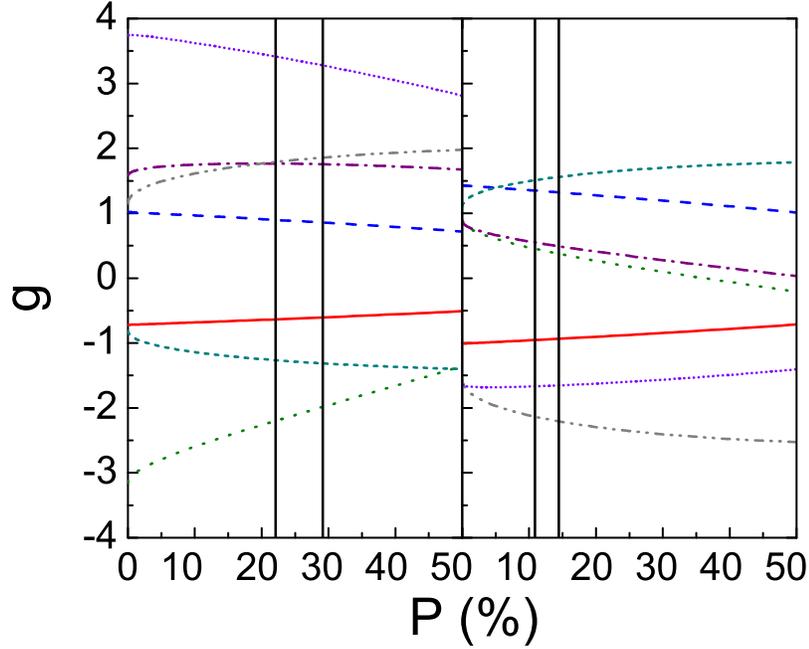}
  \caption{Coupling constants $g_{S_{11}MB}$ for $S_{11}(1535)$ (left panel) and $S_{11}(1650)$ (right panel)
  as a function of five-quark components probabilities at $\theta_{S} = 28^\circ$, for meson-baryon sets: 
  $\pi^{0}p$ (full), $\pi^{+}n$ (dashed), $\eta p$ (dotted), $K^{+}\Lambda$ (dash-dotted), 
  $K^{0}\Sigma^{+}$ (dash-dot-dotted), $K^{+}\Sigma^{0}$ (short dashed), $\eta^{\prime}p$
  (short dotted). The vertical bands correspond to the probability ranges determined in this work.}
\label{cc}
\end{figure}

The obtained model is used to put forward predictions for $S_{11}$-meson-baryon coupling constants,  
Fig.~\ref{cc}, for the following meson-baryon sets:
$\pi^{0}p$, $\pi^{+}n$, $\eta p$, $K^{+}\Lambda$, $K^{0}\Sigma^{+}$, $K^{+}\Sigma^{0}$, $\eta^{\prime}p$.
In Fig.~\ref{cc}, the couplings values at null five-quark probabilities correspond to pure qqq configuration
within broken $SU(6)\otimes O(3)$ symmetry. 

The coupling $g_{S_{11}(1535) \eta N}$ turns out to be the most sensitive one to
the five-quark components. Significant effects are also anticipated for $g_{S_{11}(1535) K^0 \Sigma^+}$ and 
$g_{S_{11}(1650) K^0 \Sigma^+}$.

\vspace{2pt}
With respect to the magnitude of couplings, we find the following orderings:

\vspace{5pt}
- for $S_{11} \equiv S_{11}(1535)$:
\begin{eqnarray}
|g_{S_{11}\pi^{0}p}| < |g_{S_{11}\pi^{+}n}| < |g_{S_{11}K^{+}\Sigma^{0}}| < |g_{S_{11}K^{+} \Lambda} |
< |g_{S_{11}K^{0}\Sigma^{+}}| < |g_{S_{11}\eta p}| < |g_{S_{11}\eta^{\prime}p}|, 
\end{eqnarray}

\vspace{5pt}
- for $S_{11} \equiv S_{11}(1650)$:
\begin{eqnarray}
|g_{S_{11}\eta p}| < |g_{S_{11}K^{+} \Lambda} | < |g_{S_{11}\pi^{0}p}| < |g_{S_{11}\pi^{+}n}| < 
|g_{S_{11}K^{+}\Sigma^{0}}| < |g_{S_{11}\eta^{\prime}p}| < |g_{S_{11}K^{0}\Sigma^{+}}|.  
\end{eqnarray}

\section{Conclusion}

The partial decay widths for the low-lying $S_{11}$ resonances are known with about
14\% acurracy for $\Gamma_{S_{11}(1535) \to \eta N}$ and $\Gamma_{S_{11}(1650) \to K^+ \Lambda}$, 
22\% for $\Gamma_{S_{11} \to \pi N}$ for both resonances, and 
95\% for $\Gamma_{S_{11}(1650) \to \eta N}$.
Previous studies~\cite{Inoue:2001ip,Penner:2002ma} have allowed to reproduce a significant 
number of those data.
Chiral constituent quark model without or with $SU(6)\otimes O(3)$ breaking effects 
lead~\cite{An:2011} to significant discrepancies with the data.

In the present work we extended the $\chi$CQM approach by complementing it with five-quark components
in the $S_{11}$ resonances wave functions. 
The phase space formed by the mixing angle and the probabilities of those components, leads
{\it simultaneously} to satisfactory results with respect to all five measured widths, within the
following ranges: $26.9^\circ \leq \theta_{S}\leq 29.8^\circ$, 
$21\% \leq P_{5q} \leq 31\% $ and 
$11\% \leq P^\prime_{5q} \leq 18\%$.
Here we have reported predictions for the coupling constants of
$\pi^{0}p$, $\pi^{+}n$, $\eta p$, $K^{+}\Lambda$, $K^{0}\Sigma^{+}$, $K^{+}\Sigma^{0}$, $\eta^{\prime}p$
pseudoscalar meson - octet baryon sets to both $S_{11}$ resonances.

A worthy to be noticed outcome of the present work is that the $S_{11}(1535) \eta N$ system turns out
to be very appealing for two reasons: i) the partial decay width $\Gamma_{S_{11}(1535) \to \eta N}$ is sizeable, 
ii) the coupling constant $g_{S_{11}(1535)\eta N}$ is rather large and sensitive enough to the higher 
Fock space components in the resonance. 
Accordingly, possible role of the baryon five-quark components in the $\eta$ meson production processes is under investigation.

\bibliographystyle{aipproc}   

\end{document}